\documentclass[sn-mathphys,Numbered]{sn-jnl}% Math and Physical Sciences Reference Style
\usepackage{graphicx}%
\usepackage{multirow}%
\usepackage{amsmath,amssymb,amsfonts}%
\usepackage{amsthm}%
\usepackage{mathrsfs}%
\usepackage[title]{appendix}%
\usepackage{xcolor}%
\usepackage{textcomp}%
\usepackage{manyfoot}%
\usepackage{booktabs}%
\usepackage{algorithm}%
\usepackage{algorithmicx}%
\usepackage{algpseudocode}%
\usepackage{listings}%
\theoremstyle{thmstyleone}%
\theoremstyle{thmstyletwo}%

\theoremstyle{thmstylethree}%

\usepackage[capitalize]{cleveref}
\usepackage{glossaries}
\glsdisablehyper     % disable hyperreferencing acronyms
\newacronym{sss}{SSC}{spin-split (or Zeeman-split) superconductor}
\newacronym{abs}{ABS}{Andreev bound state}
\begin{document}

\title[P-wave pairing near a spin-split Josephson junction]{P-wave pairing near a spin-split Josephson junction}


\author*[1]{\fnm{Rub\'en} \sur{Seoane Souto}}\email{ruben.seoane@csic.es}

\author[2]{\fnm{Dushko} \sur{Kuzmanovski}}

\author[3]{\fnm{Ignacio} \sur{Sardinero}}

\author[3]{\fnm{Pablo} \sur{Burset}}

\author[2,4]{\fnm{Alexander V.} \sur{Balatsky}}

\affil[1]{Instituto de Ciencia de Materiales de Madrid (ICMM-CSIC), Sor Juana In\'es de la Cruz 3, 28049 Madrid, Spain}

\affil[2]{Nordita, KTH Royal Institute of Technology and Stockholm University Hannes Alfv\'{e}ns v\"{a}g 12, SE-106 91 Stockholm, Sweden}

\affil[3]{Department of Theoretical Condensed Matter Physics,
Condensed Matter Physics Center (IFIMAC) and Instituto Nicol\'as Cabrera, Universidad Aut\'onoma de Madrid, 28049 Madrid, Spain}

\affil[4]{Department of Physics, University of Connecticut, Storrs, Connecticut 06269, USA}

\abstract{Superconductivity and magnetism are competing effects that can coexist in certain regimes. Their co-existence leads to unexpected new behaviors that include the onset of exotic electron pair mechanisms and topological phases. In this work, we study the properties of a Josephson junction between two spin-split superconductors. The spin-splitting in the superconductors can arise from either the coupling to a ferromagnetic material or an external magnetic field. The properties of the junction are dominated by the Andreev bound states that are also split. One of these states can cross the superconductor's Fermi level, leading to a ground state transition characterized by a suppressed supercurrent. We interpret the supercurrent blockade as coming from a dominance of p-wave pairing close to the junction, where electrons at both sides of the junction pair. To support this interpretation, we analyze the different pairing channels and show that p-wave pairing is favored in the case where the magnetization of the two superconductors is parallel and suppressed in the anti-parallel case. We also analyze the noise spectrum that shows signatures of the ground state transition in the form of an elevated zero-frequency noise.}




\maketitle

\section{Introduction}\label{sec1}

This manuscript is intended for the issue dedicated to Alexander F. Andreev. We first would like to  celebrate the remarkable accomplishments of A.~F. Andreev as a scientist, scientific leader and very open and engaging person. Some of us had a chance to meet and interact with A.~F. Andreev. One feature of Andreev's way to approach physics was the fearless embrace of bold ideas wherever the conclusions might lead. We had the chance to observe Andreev asking deep yet simple questions ranging from the one of the first work on spin nematic~\cite{Andreev1984}, work on coherent mixed parity states in mesoscopic conductors~\cite{Andreev_PS2002}, time translation violation~\cite{Andreev1996} to the nature of supersolid states in He~\cite{Andreev1969}. It was always interesting and revealing to observe how A.~F. handled the intellectual back and forth in the discussions where concepts and ideas were not always accepted right away. His approach to physics left a deep impression on the younger generation of physicists in the orbit of Kapitza and Landau schools. The shining example of his logic and intuition is the work on Josephson effect and the role of single particle states that are formed in the Josephson junctions. Extensions of these ideas to spin polarized contacts is the focus of this paper. 

%Superconductivity has been a central topic in the field of condensed matter physics, revealing phenomena that helped developing our understanding of quantum mechanics and electron dynamics in materials. Among these phenomena, Andreev reflections and Andreev bound states (ABSs) stand out for their unique contributions to our understanding of superconductor interfaces.

The key concept of Andreev reflection was first introduced by A.~F. Andreev in 1964~\cite{Andreev1964}. At a time when the BCS theory (developed by J. Bardeen, L. Cooper, and R. Schrieffer) had just provided a microscopic explanation for superconductivity, Andreev's work offered a novel insight into how superconducting and normal conductive materials interact at their interface. Andreev discovered that, at the interface between a superconductor and a normal conductor, an incident electron from the normal side could be reflected as a hole, while a Cooper pair is added to the superconductor. This reflection process, now known as Andreev reflection, revealed an entirely new mechanism of charge transfer across superconductor-normal metal interfaces; fundamentally distinct from ordinary electron scattering.

Building on the understanding of Andreev reflections, the concept of Andreev bound states emerged as a direct consequence of these reflections at superconductor interfaces~\cite{Andreev1966}. Under certain conditions, such as in the presence of a weak link or a constriction in the superconductor, Andreev reflection processes can lead to the formation of localized energy states, known as Andreev bound states (ABSs). These states, characterized by an energy within the superconducting gap, dominate the low-energy properties of superconducting junctions~\cite{Sauls_review}.

%Odd frequencyOdd freq. in He \cite{Berezinskii_JETP1974}; odd freq. in superconductors \cite{Balatsky_PRB1992}. Long range proximity effect: \cite{Bergeret_2001PRL}, review:\cite{Bergeret_2005RMP}, experiment \cite{DiBernardo_2015PRX}.

The Josephson effect~\cite{Josephson1962, Anderson_PRL1963} is the flow of dissipationless supercurrent through a tunnel junction between two superconductors with different superconducting phases. The current-phase relation for highly transparent junctions is mainly determined by the energy dispersion of ABSs as a function of the phase~\cite{Golubov2004}. Under certain conditions, these states can cross the superconductor's Fermi level, changing the ground state and reversing its current contribution, the so-called $\pi$ phase. 
% In junctions made out of Zeeman-split superconducting leads, spin-degenerate ABSs split can cross zero-energy for a phase-bias smaller (larger) than $\pi$.

Recently, there has been a revived interest in heterostructures using spin-split superconducting leads as a prospective application in superconducting spintronics~\cite{Linder_NatPhys2015, Eschrig_RPP2015}, thermoelectricity~\cite{Bergeret2018}, and to engineer materials with topological properties~\cite{Vaitiekenas_NatPhys2021,flensberg2021engineered}.

\begin{figure}[t]%
\centering
\includegraphics[width=0.9\textwidth]{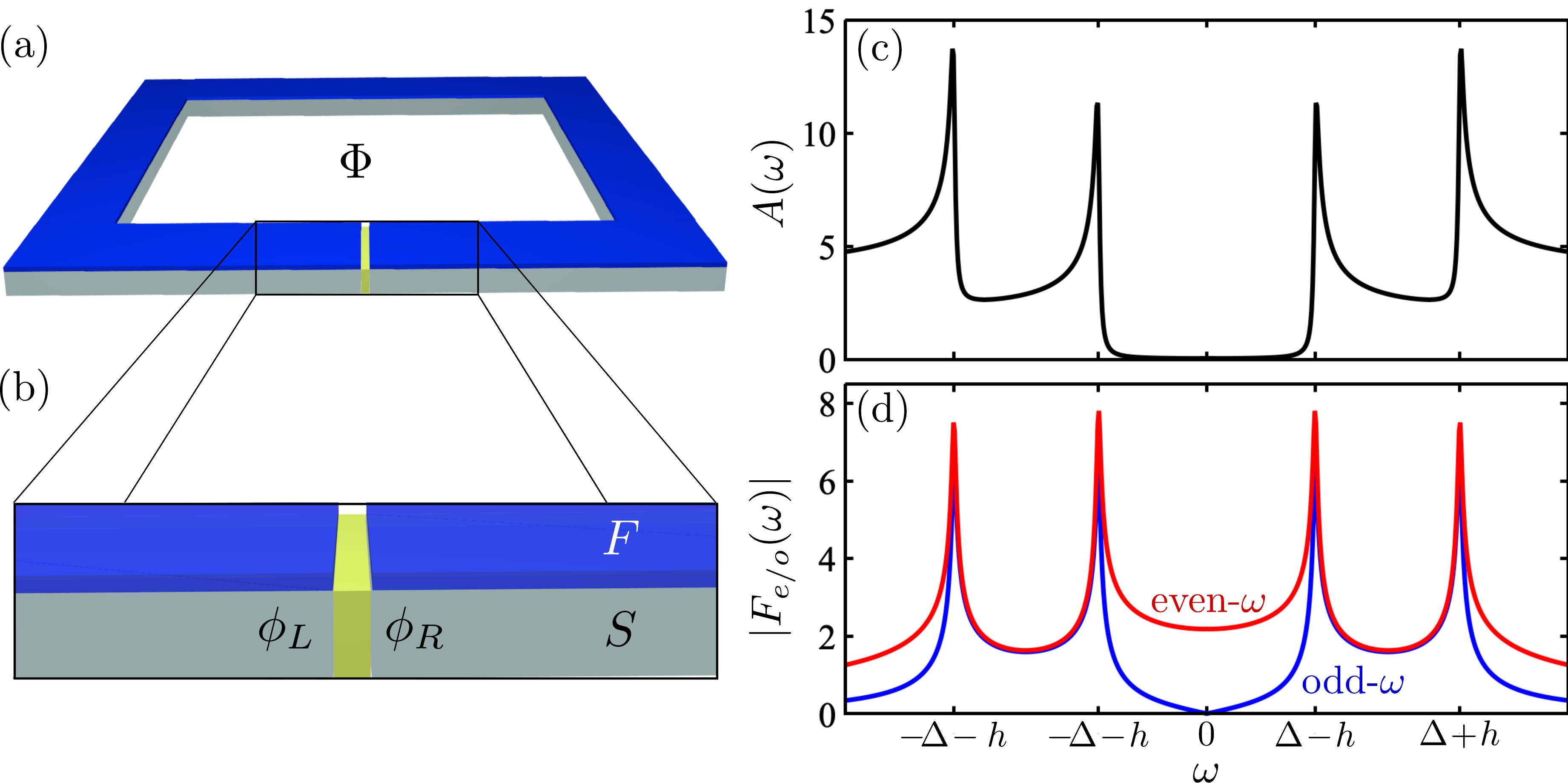}
\caption{(a) Sketch of the setup consisting of loop coupling two spin-split superconductors. (b) Junction interface. (c) Density of states at the interface. (d) Even-$\omega$ (red) and odd-$\omega$ (blue) pair amplitudes. }\label{fig1}
\end{figure}

In this work, we consider a quantum-point-contact Josephson junction (JJ) with spin-split superconducting leads, see Fig.~\ref{fig1}. The spin splitting is achieved via proximity coupling to ferromagnetic insulators with large intrinsic Zeeman exchange fields. The phase-biased JJ shows two pairs of spin-split in-gap ABSs. Two of them can cross zero energy at finite phase difference.

We find that the system can host peculiar pair amplitudes. First, the induced exchange field generates a spin-triplet, on-site, odd-frequency pairing due to spin-rotation symmetry breaking. Second, translation-symmetry breaking generates a spin-triplet, non-local p-wave pairing across the JJ. These unconventional pair amplitudes follow the SPOT rule~\cite{Linder_RMP2019} and have different frequency behavior.

The zero-frequency-crossing phase interval delimits a flat zero-current branch in the current-phase relation. We attribute this suppression to a compensating contribution of the two ABSs with opposite spin. This region is characterized by enhanced zero-frequency noise and the dominance of p-wave pairing close to the junction.

The rest of the article is organized as follows. In Sec.~\ref{sec3}, we introduce the model and the formalism. The main results are presented in Sec.~\ref{sec:results}. We present the density of states and the current in Sec.~\ref{sec:current}. We correlate the observed sharp features in the current with the onset of p-wave pairing at the junction in Sec.~\ref{sec:p-wave}. The dependence on the angle magnetization is discussed in Sec.~\ref{sec:Bangle}. In Sec.~\ref{sec:noise}, we analyze the current noise. Finally, we present the conclusions of our work in Sec.~\ref{sec4}.

%%%%%%%%%%%%%%%%%%%%%%
%%%%%%%%%%%%%%%%%%%%%%

\section{Model and formalism}\label{sec3}

%%%%%%%%%%%%%%%%%%%%%%
We consider a Josephson junction consisting of two one-dimensional semi-infinite superconductors coupled by a weak link, see \cref{fig1}. 
The system is described by the Hamiltonian~\cite{Lu_2020,Yang2023}
\begin{equation}
    \hat{\mathcal{H}}=\hat{\mathcal{H}}_{L}+\hat{\mathcal{H}}_{R}+\hat{\mathcal{H}}_{T}\,,
\end{equation}
with $\hat{\mathcal{H}}_{L}$ and $\hat{\mathcal{H}}_{R}$ representing the left and right bulk superconductors respectively
\begin{equation}
	\mathcal{\hat{H}}_{x=L,R}=\frac{1}{2}\sum_{k}\hat{\Psi}_{x}^{\dagger }(k)
	\begin{pmatrix}
		\hat{H}_{x}(k) & \hat{\Delta}_x(k) \\
		\hat{\Delta}^{\dag}_x(k) & -\hat{H}_{x}^{T}(-k)
	\end{pmatrix}
	\hat{\Psi}_{x} (k).  \label{eq:hamil-LR}
\end{equation}
Here, \cref{eq:hamil-LR} is written in Nambu (particle-hole) and spin space with basis $\hat{\Psi}_{x}(k)= [ \mathbf{c}_{x,\uparrow }(k),\mathbf{c}_{x,\downarrow }(k),\mathbf{c}_{x,\uparrow }^\dagger(-k), \mathbf{c}_{x,\downarrow }^\dagger(-k)] ^{T}$, where $\mathbf{c}_{x,\sigma}(k)$ [$\mathbf{c}_{x,\sigma}^\dagger(k)$] are creation [annihilation] operators for electrons with spin $\sigma$ and momentum $k$ at the lead $x=L,R$. 
The non-interacting Hamiltonian on each lead adopts the form
\begin{equation}
	\hat{H}_{x}(k) =\left( \frac{\hbar^2k^{2}}{2m} - \mu \right) \hat{\sigma}_{0} - g\mu_{B} \hat{\mathbf{\sigma}} \cdot \mathbf{M}_{x} ,
\end{equation}
with the Pauli matrices $\hat{\sigma}_{0}$ and $\hat{\mathbf{\sigma}}=(\hat{\sigma}_x,\hat{\sigma}_y,\hat{\sigma}_z)$ acting on spin space and $g$, $\mu _{B}$, and $\mathbf{M}_{L,R}$ respectively being the effective Land\'e $g$-factor, Bohr magneton, and induced exchange field. We consider a symmetric junction where both superconductors have the same effective mass $m$ and spin-singlet s-wave pair potential $\hat{\Delta} (k) = i\hat{\sigma}_{y}\Delta $, with $\Delta>0$. 

We assume that the spin-splitting fields are induced on each superconductor by proximity to a ferromagnetic material, resulting in the Zeeman fields $\mathbf{M}_{L,R}$, see \cref{fig1}. In principle, these spin-fields can have an arbitrary axis. For simplicity, we assume that both Zeeman fields lie in the $x-z$ plane and impose that $\mathbf{M}_{L} = M_L\mathbf{\hat{z}}$ and $\mathbf{M}_{R} = M_R \left( \sin\alpha, 0, \cos\alpha\right) $,
where $\alpha$ is the relative angle between the magnetization of both leads and $h_x=g\mu_B|M_x|$ represents the magnitude of the proximity-induced spin fields. We only consider fields with magnitude within the Chandrasekhar-Clogston limit: $h_x<\Delta/\sqrt{2}$~\cite{Chandrasekhar_1962,Clogston_1962}. 
If the relative orientation $\alpha$ is maintained, the direction of the spin-splitting fields can be trivially changed by a rotation without affecting our results. 

Finally, the tunneling Hamiltonian is described by
\begin{equation}\label{eq:hamil-T}
	\hat{\mathcal{H}}_{T}(t) = \frac{1}{2}\sum_{k,k'}\left[\hat{\Psi}^{\dagger}_{L }(k) \hat{T}_{LR} \hat{\Psi}_{R}(k') + \mathrm{h.c.}\right] ,
\end{equation}
where the Pauli matrices $\hat{\tau}_{0,x,y,z}$ act in Nambu space and 
\begin{equation}
    \hat{T}_{LR} = t \delta_{kk'} \hat{\tau}_{z} \mathrm{e}^{i \phi \hat{\tau}_{z}/2}\,,
\end{equation}
with $\phi$ being the phase difference between superconductors. \Cref{eq:hamil-T} assumes that there is no mixing of channels $k$ at the junction. We then follow the quasi-classical approximation and average the summation over channels $k$~\cite{Lu_2016,Burset_2017,Lu_2020}. 

The properties of the junction can be described using the retarded (r) and advanced (a) Green functions, 
\begin{equation}\label{eq:GF-def}
    \hat{G}_{xy}^{r,a}\left( \omega \right)=-i\langle \mathcal{T} \left[\hat{\Psi}_{x} (t) \hat{\Psi}^{\dagger}_{y}(t') \right] \rangle, 
\end{equation}
where $y=L,R$ and $\mathcal{T}$ is the time ordering operator. To calculate the full Green function, we use the boundary Green functions from the leads at the edge of each semi-infinite superconductor~\cite{Cuevas_1996}:
\begin{equation}
	\mathcal{\hat{G}}_{x}^{r,a}= 
	\begin{pmatrix}
		g_{x,\uparrow }^{r,a} & 0 & 0 & f_{x,\uparrow }^{r,a} \\
		0 & g_{x, \downarrow }^{r,a} & f_{x,\downarrow }^{r,a} & 0 \\
		0 & f_{x, \downarrow }^{r,a} & g_{x, \downarrow }^{r,a} & 0 \\
		f_{x, \uparrow }^{r,a} & 0 & 0 & g_{x, \uparrow }^{r,a}
	\end{pmatrix} ,
\end{equation}
where
\begin{subequations}\label{eq:BCS-GFs}
\begin{align}
	g_{x, \sigma }^{r,a}\left( \omega \right) ={}& 
	\frac{-\left( \omega + s_{\sigma } h_x \pm i\eta \right) }{ W_x \sqrt{\Delta^{2} - \left( \omega + s_{\sigma } h_x \pm i \eta \right)^2 }},
	\\
	f_{x, \sigma }^{r,a}\left( \omega \right) ={}& 
	\frac{\Delta}{W_x \sqrt{\Delta^{2} - \left( \omega + s_{\sigma } h_x \pm i \eta \right)^2 }},
\end{align}
\end{subequations}
with $s_\sigma= +(-)1$ for the spin projection $\sigma=\uparrow,\downarrow$ that is parallel (anti-parallel) to the direction of the exchange field and $W_L=W_R\equiv W$ being the superconductor's bandwidth~\cite{Cuevas_2001}. 

To take into account the relative orientations of the spin fields on each superconductor we define the rotation $\hat{U}_\alpha \!=\! e^{-i\hat{\sigma}_{y}\alpha/2} \hat{\tau}_{0}$ and impose~\cite{Lu_2020}
\begin{equation}\label{eq:rotated-GF}
	\mathcal{\hat{G}}_{R}^{r,a}= \hat{U}_\alpha \mathcal{\hat{G}}_{L}^{r,a} \hat{U}_\alpha^\dagger . 
\end{equation}

When the Josephson junction is phase-biased, that is, the two superconductors are at thermal equilibrium and there is no applied voltage across the junction, the supercurrent can be expressed as~\cite{Kashuba_2017}
\begin{equation}\label{eq:current}
	I(\phi)= \frac{e}{2\hbar} \int \frac{\mathrm{d}\omega}{2\pi} \tanh\left( \frac{\omega}{2k_BT} \right) 
	\mathrm{Tr} \left\{ 
		\hat{\tau}_z \left[ \hat{\Sigma}_R^r(\omega) \hat{G}_{LL}^r(\omega) - \hat{\Sigma}_R^a(\omega) \hat{G}_{LL}^a(\omega) \right] 
	\right\} , 
\end{equation}
with temperature $T$ and $k_B$ being the Boltzmann constant. 
The Green functions in \cref{eq:current} are obtained using Dyson equation as 
\begin{equation}\label{eq:GF}
	\hat{G}_{LL}^{r,a}(\omega) = \left(
	[\hat{\mathcal{G}}_{L}^{r,a}(\omega)] ^{-1} - \hat{\Sigma}_{R}^{r,a}
	\right)^{-1} ,
\end{equation}
with 
\begin{equation}
	\hat{\Sigma}_{R}^{r,a}(\omega) = \hat{T}_{LR} \hat{\mathcal{G}}_{R}^{r,a}(\omega) \hat{T}_{LR}^\dagger ,
\end{equation}
where we have averaged over all channels. We can analogously define $\hat{G}_{RR}^{r,a}(\omega)$ and $\hat{\Sigma}_{L}^{r,a}(\omega) $, including the rotation given in \cref{eq:rotated-GF}. We can also obtain the non-local Green functions using
\begin{equation}
	\hat{G}_{x\bar{x}}^{r,a}(\omega)=\hat{\mathcal{G}}_{xx}^{r,a}(\omega) \hat{T}_{x\bar{x}}\hat{G}_{x\bar{x}}^{r,a}(\omega) ,
\end{equation}
with $\bar{x}=R,L$ when $x=L,R$. 
Finally, the Keldysh components of the Green functions in equilibrium can be obtained from the retarded and advanced components as 
\begin{equation}
	\hat{G}_{xy}^{+-}(\omega)=n_F(\omega)\left[\hat{G}_{xy}^{a}(\omega)-\hat{G}_{xy}^{r}(\omega)\right]\,,
\end{equation}
with $n_F$ being the Fermi distribution function. 

The fully coupled Green function in \cref{eq:GF} allows us to compute the density of states at the interface as
\begin{equation}\label{eq:int-DOS}
	A(\omega) = -\frac{1}{\pi} \mathrm{Im} \left\{ \mathrm{Tr} \left[\hat{G}_{LL}^{r}(\omega) \right] \right\} .
\end{equation}

The current in the phase-biased Josephson junction we consider is mostly carried by \glspl{abs}. 
For parallel ($\alpha=0$) splitting in the two superconductors with equal magnitude $h_L=h_R\equiv h$ the \glspl{abs} appear at energies
\begin{equation}
	\varepsilon_\sigma=\Delta\sqrt{1-\tau\sin^2(\phi/2)}+s_\sigma h\,,
	\label{Eq:ABSs}
\end{equation}
where $\tau$ is the junction transmission in the normal state ($\Delta=0$). 

In this work, we analyze the pairing amplitude contained in the Nambu off-diagonal Green functions $F_{\uparrow\downarrow,xy}(\omega)=[\hat{G}^{+-}_{xy}(\omega)]_{14}$ and $F_{\downarrow\uparrow,xy}(\omega)=[\hat{G}^{+-}_{xy}(\omega)]_{23}$. We use boundary Green functions for each superconductor, therefore computing the pairing amplitude close to the junction. The anomalous Green function $F$ changes sign under the exchange of spin, position, and time variables. If we denote these exchange operations as $S$, $P^*$, and $T^*$, it means that $S\,P^*T^*=-1$. Therefore, there are 4 possible pairing amplitudes that can appear in the system: $F^{-++}_{xy}(\omega)$ (the equivalent to BCS pairing for $x=y$), $F^{+-+}_{xy}(\omega)$ (p-wave pairing), $F^{++-}_{xy}(\omega)$, and $F^{---}_{xy}(\omega)$. The latter two are the odd-frequency components. Here, we have used the notation $F^{spt}$ with $s$, $p$, $t$ being the eigenvalues of the argument exchange operations, $S$, $P^*$, and $T^*$. 

To get insight about the pairing mechanisms at the junction, we symmetrize the anomalous Green function $F$ into components with a well-defined sign after the exchange of one of the coordinates. For the local pairing with $x=y$ and, consequently, $p=+1$, we only have two possibilities
% local: Even Singlet (-++) and odd-triplet (++-)
\begin{equation}\label{eq:anom-GF0}
    F^{\mp + \pm}_{xx}(\omega)=\left[F_{\uparrow\downarrow,xx}(\omega) \pm F_{\downarrow\uparrow,xx}(\omega)\right]/2 .
\end{equation}
We can also analyze the non-local pair amplitude: Cooper pairs formed by electrons belonging to different sides of the junction. In this case, we have four possible components, two for spin-singlet
% nonlocal, all cases: -++, ---, ++-, +-+
%\begin{equation}
%    F^{spt}_{x\bar{x}}(\omega)=\left\{F_{\uparrow\downarrow,x\bar{x}}(\omega)+sF_{\downarrow\uparrow,x\bar{x}}(\omega)+p\left[F_{\uparrow\downarrow,\bar{x}x}(\omega)+sF_{\downarrow\uparrow,\bar{x}x}(\omega)\right]\right\}/4 ,
%\end{equation}
%with $t=-s\,p$ and $\bar{x}$ denotes the opposite side of the junction with respect to $x$.
\begin{equation}\label{eq:anom-GF1}
    F^{-\pm\pm}_{x\bar{x}}(\omega)=\left\{
    F_{\uparrow\downarrow,x\bar{x}}(\omega) \pm
    F_{\downarrow\uparrow,x\bar{x}}(\omega) -
    F_{\uparrow\downarrow,\bar{x}x}(\omega) \mp
    F_{\downarrow\uparrow,\bar{x}x}(\omega) \right\}/4 ,
\end{equation}
and two for triplet states: 
\begin{equation}\label{eq:anom-GF2}
    F^{+\mp\pm}_{x\bar{x}}(\omega)=\left\{
    F_{\uparrow\downarrow,x\bar{x}}(\omega) \mp
    F_{\downarrow\uparrow,x\bar{x}}(\omega) +
    F_{\uparrow\downarrow,\bar{x}x}(\omega) \mp
    F_{\downarrow\uparrow,\bar{x}x}(\omega) \right\}/4 . 
\end{equation}

\section{Results}\label{sec:results}

The system we study is introduced in Fig.~\ref{fig1}(a,b), where two spin-split superconductors couple via a weak link. In the following, we consider that the junction between the two superconductors has a single channel with relatively high transmission. Similar results will hold for a multichannel junction as long as there is no coupling between the channels. 

Far away from the junction, the spin-split superconductors feature a split density of states, preserving the gap $\Delta$ for each of the two spin-spices. This gap is however not centered around the superconductor's Fermi level, but displaced by the exchange field, $\pm h$, depending on the spin direction, see Fig~\ref{fig1}(c). 
% Spin-split superconductors can be realized by applying external magnetic field. However, orbital effects are detrimental and suppress superconductivity. For this reason, magnetic materials are a more controlled way to reach sizeable spin-splitting without suppressing superconductivity.
We consider the superconductors to be native BCS: formed by spin singlet s-wave pairs, see \cref{eq:hamil-LR}. The spin-splitting forces a conversion between spin-singlet and spin-triplet pairs at the bulk of the superconductor, as shown in Fig~\ref{fig1}(d). These spin-triplet pairs are odd in frequency~\cite{Linder2015_SciRep2015} to preserve fermion anti-commutation relations~\cite{Linder_RMP2019}. 

\subsection{Junction properties}\label{sec:current}

The physical properties of the junction are dominated by the Andreev reflections: conversions of electrons that retro-reflect into holes leading to the transfer of Cooper pairs. These Andreev reflections generate Andreev bound states that are localized at the interface between the two superconductors. The expression for the energy of the Andreev bound states is analytic and given in Eq.~\eqref{Eq:ABSs}, which predicts a minimal gap between the Andreev states of $\delta \varepsilon=2\Delta \sqrt{1-\tau}$ at $\phi=\pi$ for $h=0$.

\begin{figure}[t]%
\centering
\includegraphics[width=0.9\textwidth]{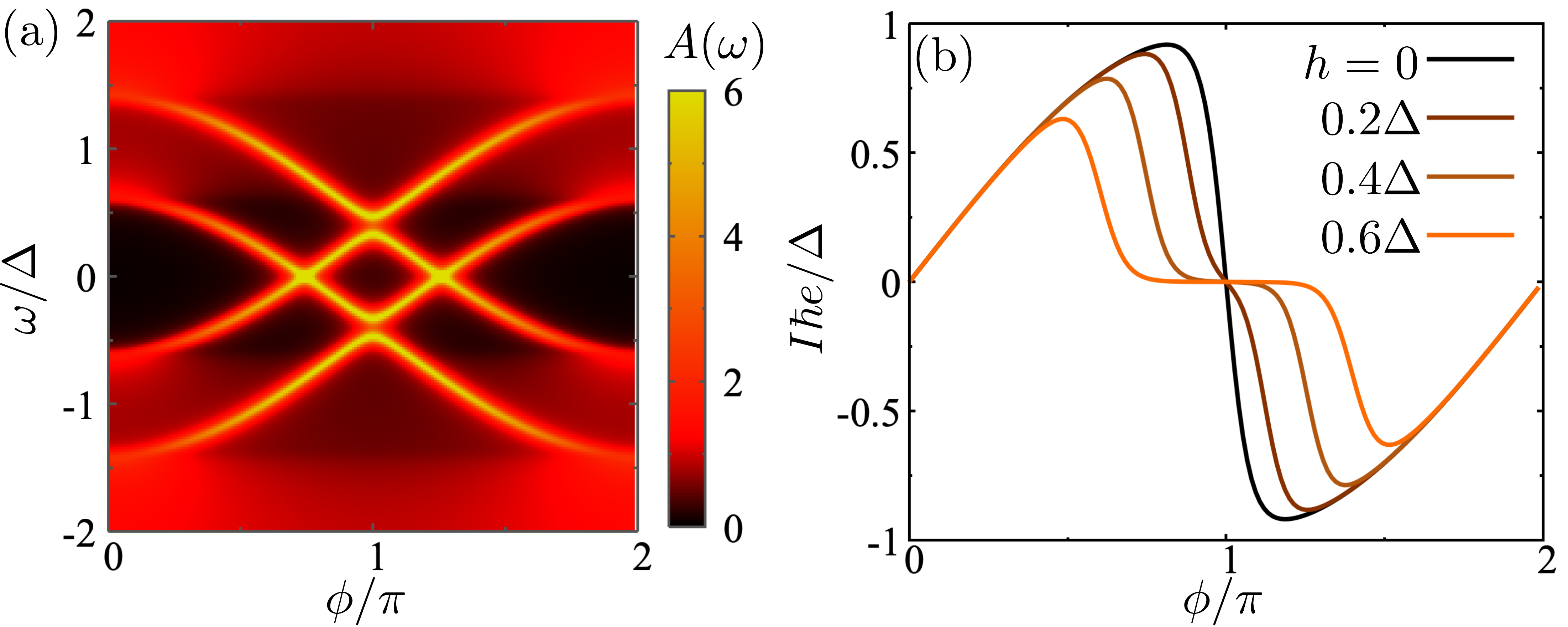}
\caption{(a) Density of states at the left lead as a function of the phase difference $\phi$. (b) Supercurrent as a function of the phase for different spin-splitting fields in the leads. The rest of parameters are $\Delta=1$, $\alpha=0$ and $t=0.93$, corresponding to $\tau=0.995$. We consider $h_L=h_R$ and take them equal to $0.4\Delta$ for panel (a). }\label{fig2}
\end{figure}

The magnetization in the superconductors induces a spin-splitting of the Andreev bound states, see Fig.~\ref{fig2}(a). For parallel and equal magnetization in the two superconductors, the Andreev states split by $\pm h$. If the exchange field is larger than the minimal gap between the states, $h>\delta \varepsilon$, two Andreev bound states with opposite spin can cross the Fermi level [Fig.~\ref{fig2}(a)]. It means that the system acquires a spin-polarization that is localized close to the junction. 

The energy of the occupied Andreev states is directly reflected in the supercurrent, that depends on the phase derivative of the Andreev states. More precisely, at zero temperature the current contribution of the Andreev states below the Fermi level is $I\propto \partial_\phi \varepsilon_\sigma(\phi)$. The magnetization only shifts the spectrum for the up and down spins without affecting to the curvature of the states. This implies that after the states cross the Fermi level their current contributions cancel, thus suppressing the supercurrent, see Fig.~\ref{fig2}(b). In the literature, this crossing is usually referred to as $0-\pi$ transition, since the current through the system can reverse sign if the junction has a finite length~\cite{Razmadze_PRB2023,Maiani_PRB2023,Suzuki_PRB2023}. In the following, we interpret the $0-\pi$ transition in terms of local and non-local Cooper pairs. 

\subsection{P-wave pairing at the junction}\label{sec:p-wave}
\begin{figure}[t]%
\centering
\includegraphics[width=0.8\textwidth]{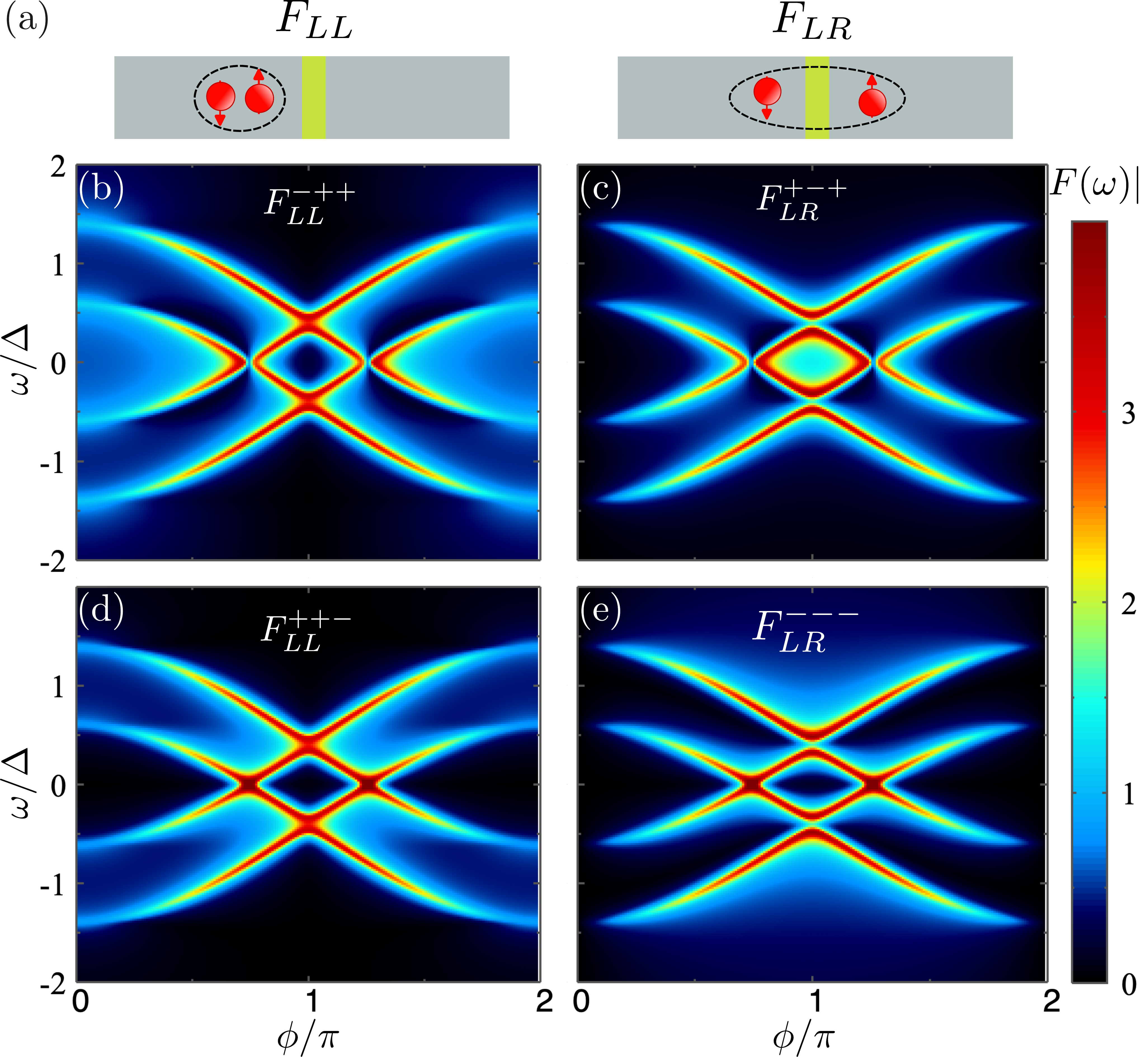}
\caption{(a) Sketches for the local and non-local Cooper pairs. (b--e) Pair amplitudes as a function of the energy and phase for even-$\omega$ (b,c) and odd-$\omega$ (d,e) components. The remaining parameters are $\Delta=1$, $h_L=h_R=0.4$, and $t=0.93$, corresponding to $\tau=0.995$.}\label{fig3}
\end{figure}

We now focus on the induced pairing at the junction, described by the anomalous Green functions in \cref{eq:anom-GF0,eq:anom-GF1,eq:anom-GF2}. For this reason, we decompose the anomalous Green function in different channels using Eqs.~\eqref{eq:anom-GF0},~\eqref{eq:anom-GF1}, and~\eqref{eq:anom-GF2}.
Figures~\ref{fig3}(b-e) show the modulus of the four components, where the first and second columns are the local and non-local pair amplitudes respectively. Top and bottom rows correspond to even-$\omega$ and odd-$\omega$ components. As shown, all the pair amplitudes peak around the energy of the Andreev bound states. This feature has been reported before in different systems, see for example Refs.~\cite{Souto_PRR2020,Kuzmanovski_PRB2020}. 

Apart from the sharp features around the Andreev states, we note that some of the pair amplitudes have a significant weight at finite frequencies. For $\phi=0$, the local singlet s-wave pairing dominates at low frequency, see Fig.~\ref{fig3}(b), while the triplet odd-$\omega$ pairing appears at energies between the two Andreev states, see Fig~\ref{fig3}(d). This is consistent with the previous results presented in Fig.~\ref{fig1}. In this situation, the left-right symmetry imposes a vanishing p-wave pair amplitude. 

A different phase between the superconductors break the left-right symmetry. In this situation, the non-local odd-parity (e.g., p-wave) pairing develops, Figs.~\ref{fig3}(c,e). The superconducting phase also modulates the energy of the ABSs, reducing their gap and, eventually inducing a crossing between them, see Fig.~\ref{fig2}(a). After the crossing, there is a suppression of the BCS (singlet s-wave) pair amplitude and an enhancement of the p-wave spin-triplet pair amplitude at low frequencies, see Fig.~\ref{fig3}(b,c). That means that the low-frequency properties of the junction are dominated by the onset of tripled p-wave pairs across the junction. These triplet pairs have zero spin-polarization in any axis. The creation of spin-polarized triplet pairs would require spin-mixing terms that can come from different magnetization angles or other spin-mixing mechanisms, like spin-orbit coupling. This fact forms the basis for proposals to engineer topological phases in macroscopic junctions~\cite{Pientka_PRX2017,Hell_PRL2017,Laeven_PRL2020,Paudel_PRB2021,Sardinero_arXiv2024}. 

\subsection{Magnetization angle}\label{sec:Bangle}

\begin{figure}[t]%
\centering
\includegraphics[width=1\textwidth]{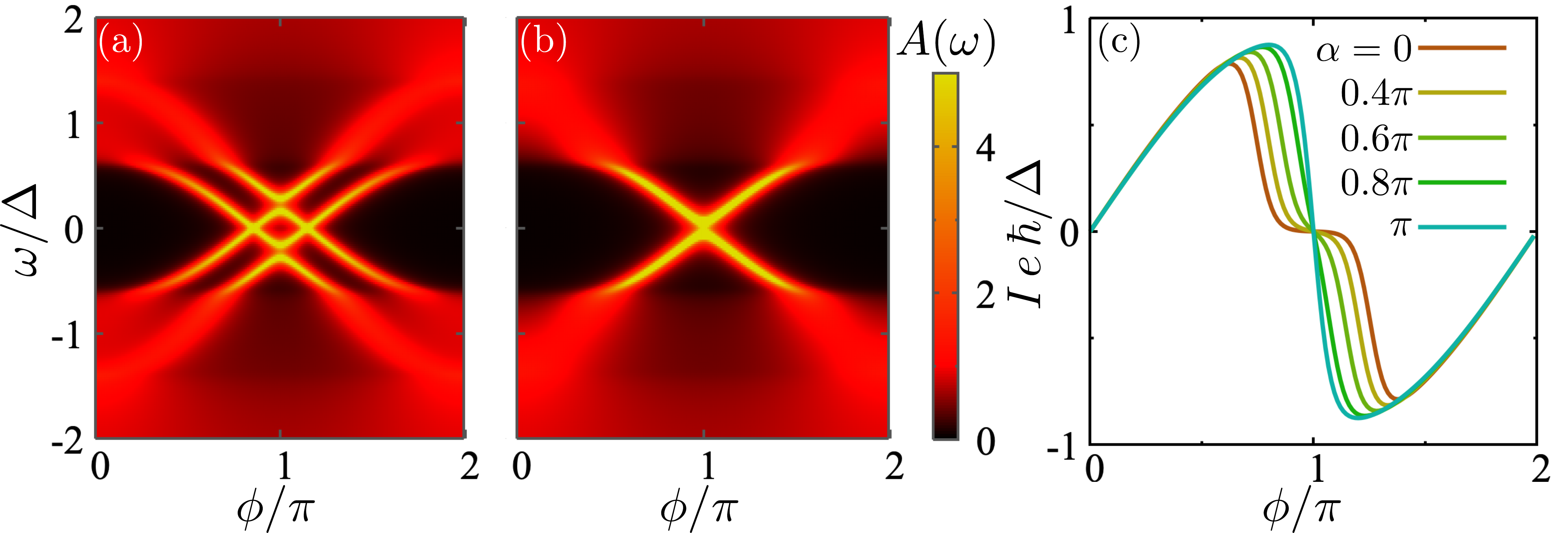}
\caption{(a,b) Density of states at the junction interface for a finite magnetization angle $\alpha/\pi=0.6$ (a) and $\alpha=\pi$ (b) between the left and right superconductors. (c) Supercurrent as a function of the phase for different values of the angle. Parameters: $|h_L|=|h_R|=0.4\Delta$, $t=0.93\Delta$, corresponding to $\tau=0.995$.}\label{fig4}
\end{figure}

We now discuss the case where the magnetization in the two superconductors is no aligned, but presents a misalignment angle $\alpha$. The misalignment of the magnetization results in the spin projection not being a good quantum number. If we take the same spin quantization axis in both leads, a tunneling electron with spin up converts into a superposition between up and down spin in the other lead. The amplitudes of both spin components depend on the projection between the two quantization axis. We note that spin-orbit coupling in the tunnel between the superconductors would have similar effects. 

Figure~\ref{fig4}(a) shows the density of states at finite magnetization angle $\alpha=0.6\pi$. In this case, the crossing between the states happens closer to the phase $\phi=\pi$, compared to the $\alpha=0$ case, see Fig.~\ref{fig2}(a) for comparison. 
We associate this behavior to a reduction of the effective exchange field in the junction. In particular, the bound states collapse for $\alpha=\pi$, becoming degenerate, see Fig.~\ref{fig4}(b). The energy spectrum in this case resembles the $h_L=h_R=0$ case but with a smaller gap, given by $\Delta-|h|$. We also note that the bound states at higher energy broaden when they enter the continuum of states of the other spin. This feature can be used to estimate the effective angle misalignment between the quantization axis in the two superconductors. 

The magnetization angle has also an influence on the supercurrent through the system. As illustrated in Fig.~\ref{fig4}(c), the supercurrent blockade state reduces when increasing $\alpha$, disappearing for $\alpha=\pi$. 

\begin{figure}[t]%
\centering
\includegraphics[width=1\textwidth]{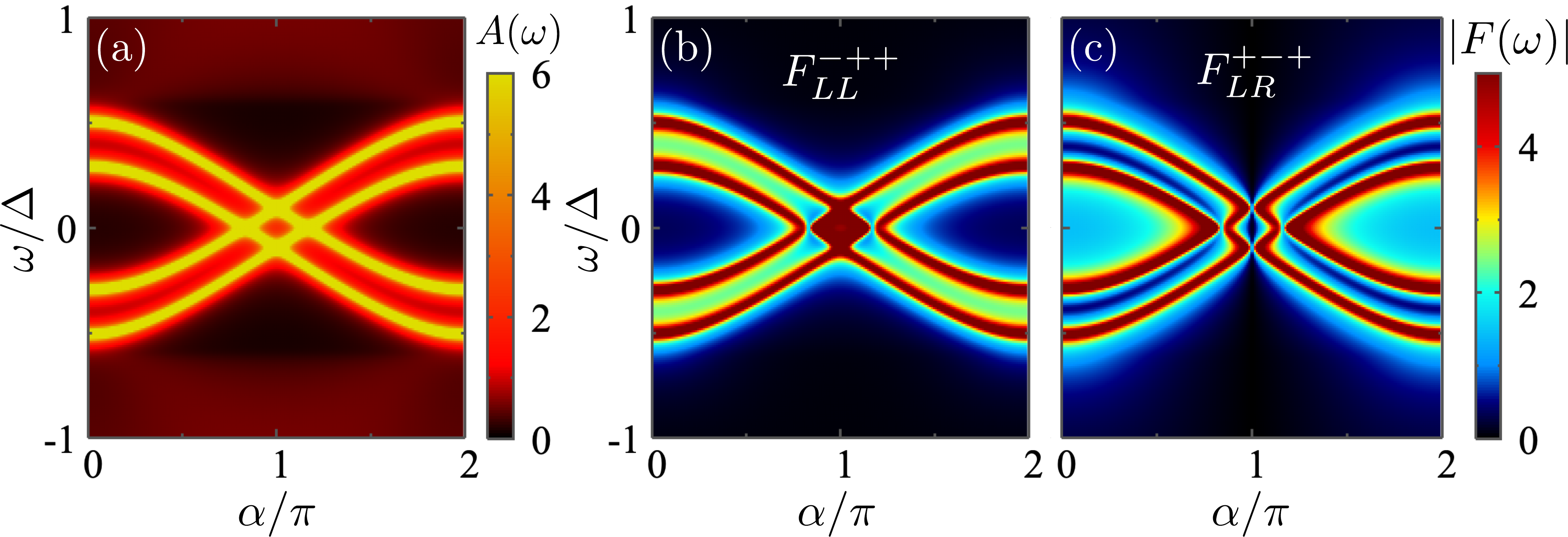}
\caption{(a) Density of states as a function of the magnetization angle of the two superconductors. (b,c) Local (BCS) and non-local (p-wave) even-$\omega$ components of the pair amplitude for $h=0.4\Delta$. The remaining parameters are $\Delta=1$, $h_L=h_R=h$, and $t=0.93\Delta$, corresponding to $\tau=0.995$.}\label{fig5}
\end{figure}

The angle has also an influence on the pairing amplitudes as presented in Fig.~\ref{fig5}. In the figure, we focus at a point where the supercurrent is suppressed ($\phi=0.9\pi$ for the same parameters as in Fig.~\ref{fig2}). For parallel magnetizations, p-wave pairing dominates while BCS s-wave is suppressed at low frequencies, as commented in Sec.~\ref{sec:p-wave} for the same parameters. As a function of the angle, two ABSs cross zero energy close to $\alpha=\pi$. The crossing has an effect on the dominant pairing amplitude at low frequencies, recovering a dominance of BCS pair amplitudes and suppressing p-wave pairing, as in the case of $h=0$. This picture is consistent with a suppression of the effective exchange field in the junction that also unblocks the supercurrent, as illustrated in Fig.~\ref{fig4}(c). 

\subsection{Current noise}\label{sec:noise}
\begin{figure}[t]%
\centering
\includegraphics[width=1\textwidth]{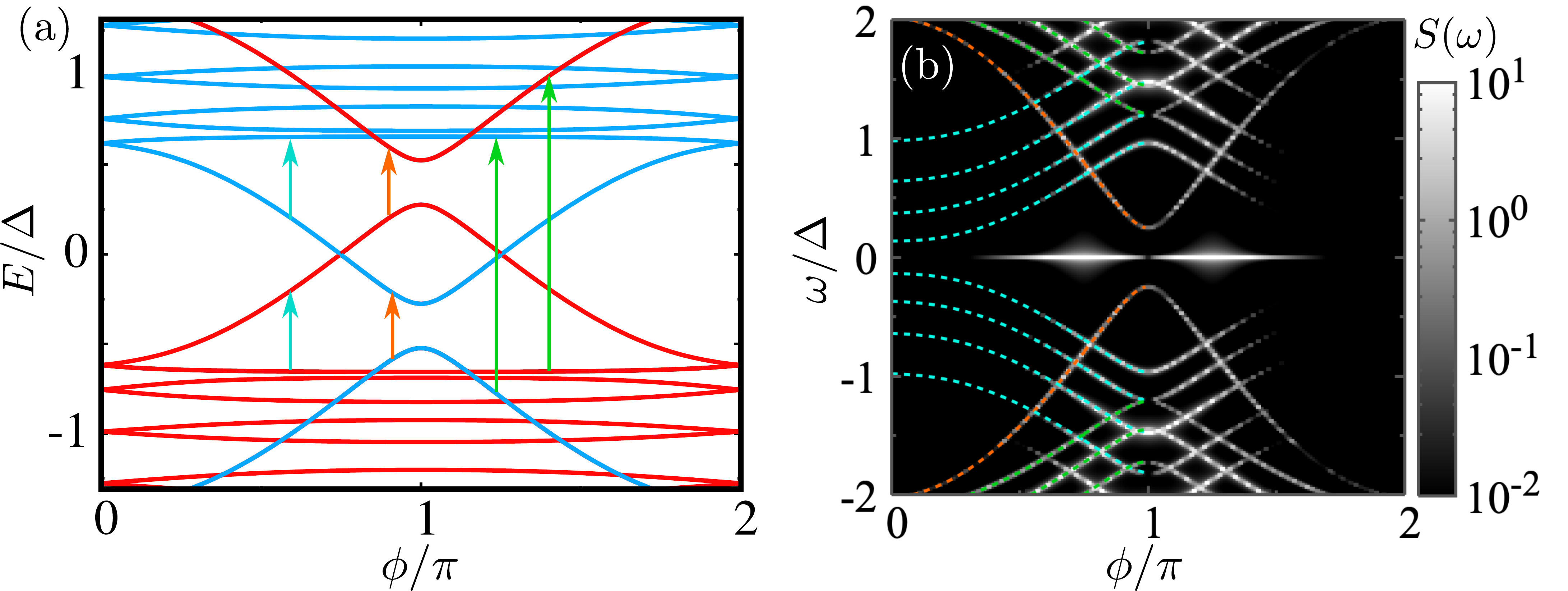}
\caption{(a) Energy spectrum of the finite-size system, discretized using a tight-binding Hamiltonian. Red and blue lines denote the two spin orientations. Due to the finite size of the system, we find a quasi-continuum of states at energies $\sim \pm|\Delta\pm h|$. (b) Noise spectrum as a function of the phase and the frequency. Dashed lines correspond to the main transitions denoted by arrows in panel (a). These transitions correspond to ABS--ABS (orange) and ABS--continuum (green and cyan). Parameters: $h_L=h_R=0.4\Delta$, $T=\Delta/20$, $\alpha=0$, and the chemical potential of the leads is $\mu = 0$, {\it i.e.}, half filling. The superconducting coherence length was chosen as $\xi/a = 8.0$. The tunneling element between the two leads is $t_1/t = 1.0$, corresponding to perfect transparency.}\label{fig6}
\end{figure}

We conclude the results section analyzing the supercurrent noise spectrum. For that, we use a tight-binding model, discretizing the left and right superconductors. We impose a superconducting phase profile that is constant in the leads and drops abruptly at the interface between the two superconductors. The energy spectrum is shown in Fig.~\ref{fig6}(a). Due to the finite size of the leads, the states in the continuum become discrete. In this subsection, we focus on the $\alpha=0$ case, where noise features are easier to understand. In this case, every line in the spectrum has a well-defined spin, denoted by the blue/red lines in Fig.~\ref{fig6}(a). 

Figure~\ref{fig6}(b) shows the noise spectrum for the same parameters as in Fig.~\ref{fig6}(a). Expressions of the current noise in the system are given in Ref.~\cite{Kuzmanovski_PRB2021}. We use dashed lines to denote the main transitions, that can occur only between states with the same spin, while spin-flip transitions are forbidden for $\alpha=0$. The orange line represents transitions between two ABSs. The green and cyan lines represent transitions between the highest- and lowest-energy Andreev states and the quasi-continuum of states. These processes are schematically represented in Fig~\ref{fig6}(a) as arrows. 

At small phases, only the states below the superconductor's Fermi level are populated. Therefore, the transitions between ABSs (orange) and ABSs to the continuum of states (green) dominate. At finite phase, two ABSs come closer, eventually crossing. Around and after the crossing, one of the highest-energy Andreev state becomes populated, contributing to the noise. This is illustrated by the additional lines appearing close to phase $\phi=\pi$, highlighted in cyan. We also note an elevated zero-frequency noise around the crossing between the states. This is due to fluctuations on the population of the Andreev states that have zero energy: a Cooper pair splitting where each of the electrons occupy one of the spin-split Andreev level. 

\section{Conclusion}\label{sec4}

In this work, we have studied the properties of spin-split Josephson junctions described by the key concept of Andreev bound states formed due to the electron-hole reflections at the interface. We have shown that the onset of a regime where supercurrent is blocked is associated with the dominance of p-wave pairing between leads, that is, non-local Cooper pairs formed by electrons from each side of the junction. The dominance of these pairs can explain the supercurrent blockade as originated from the difference between the order parameter at the bulk of each superconductor and induced pairing at the Josephson junction interface. The formation of these pairs is sensitive to the magnetization angle between the two superconductors: parallel magentizations are favorable, while the pairs are suppressed for anti-parallel orientations. We also analyze the supercurrent noise, identifying the main contributions. We show that the zero frequency noise peaks at the transition between the transmissive and blocked regimes. 

\backmatter

\bmhead{Acknowledgments}

We acknowledge Olli Mansikkamäki for interesting discussions. We acknowledge support from the Spanish CM ``Talento Program'' (project Nos. 2023-5A/IND-28927, 2022-T1/IND-24070, and 2019-T1/IND-14088), the Spanish Ministry of Science, innovation, and Universities through Grants PID2022-140552NA-I00, PID2020-117992GA-I00 and CNS2022-135950, Knut and Alice Wallenberg Foundation KAW 2019.0068, European Research Council under the European Union Seventh Framework ERS-2018-SYG 810451 HERO and University of Connecticut.

\section*{Declarations}

%{\bf Funding.} We acknowledge funding from...\\

{\bf Authors' contribution.} RSS did the transport calculations with the help of PB and the input of DK and IS. Noise calculations were performed by DK. The manuscript was written by RSS, PB, AVB, and DK with input from all the coauthors.\\

{\bf Conflict of interest. } Authors declare no conflict of interest.\\

\bibliography{bibliography}% common bib file

\end{document}